\documentclass[aps, pra, reprint, superscriptaddress, longbibliography]{revtex4-1}

\usepackage[usenames]{color}
\definecolor{darkblue}{rgb}{0,0.08,0.45}
\usepackage[bookmarksopen=true,bookmarksopenlevel=2,colorlinks=true,linkcolor=darkblue,anchorcolor=darkblue,citecolor=darkblue,urlcolor=darkblue,pdfstartview=FitH]{hyperref}
\usepackage{amsmath}
\usepackage{graphicx}
\usepackage{booktabs}

\begin{document}

\title{Shape and symmetry determine two-dimensional melting transitions of hard regular polygons}
\author{Joshua A. Anderson}
\affiliation{Department of Chemical Engineering, University of Michigan, Ann Arbor, MI 48109, USA}
\author{James Antonaglia}
\affiliation{Department of Physics, University of Michigan, Ann Arbor, MI 48109, USA}
\author{Jaime A. Millan}
\affiliation{Department of Materials Science and Engineering, University of Michigan, Ann Arbor, MI 48109, USA}
\author{Michael Engel}
\affiliation{Department of Chemical Engineering, University of Michigan, Ann Arbor, MI 48109, USA}
\affiliation{Institute for Multiscale Simulation, Friedrich-Alexander-Universit\"at Erlangen-N\"urnberg, 91058 Erlangen, Germany}
\author{Sharon C. Glotzer}
\email{sglotzer@umich.edu}
\affiliation{Department of Chemical Engineering, University of Michigan, Ann Arbor, MI 48109, USA}
\affiliation{Department of Physics, University of Michigan, Ann Arbor, MI 48109, USA}
\affiliation{Department of Materials Science and Engineering, University of Michigan, Ann Arbor, MI 48109, USA}
\affiliation{Biointerfaces Institute, University of Michigan, Ann Arbor, MI 48109, USA}
\begin{abstract}

The melting transition of two-dimensional (2D) systems is a fundamental problem in condensed matter and statistical physics that has advanced significantly through the application of computational resources and algorithms. 2D systems present the opportunity for novel phases and phase transition scenarios not observed in 3D systems, but these phases depend sensitively on the system and thus predicting how any given 2D system will behave remains a challenge. Here we report a comprehensive simulation study of the phase behavior near the melting transition of all hard regular polygons with $3\leq n\leq 14$ vertices using massively parallel Monte Carlo simulations of up to one million particles. By investigating this family of shapes, we show that the melting transition depends upon both particle shape and symmetry considerations, which together can predict which of three different melting scenarios will occur for a given $n$.  We show that systems of polygons with as few as seven edges behave like hard disks; they melt continuously from a solid to a hexatic fluid and then undergo a first-order transition from the hexatic phase to the fluid phase. We show that this behavior, which holds for all $7\leq n\leq 14$, arises from weak entropic forces among the particles. Strong directional entropic forces align polygons with fewer than seven edges and impose local order in the fluid. These forces can enhance or suppress the discontinuous character of the transition depending on whether the local order in the fluid is compatible with the local order in the solid. As a result, systems of triangles, squares, and hexagons exhibit a KTHNY-type continuous transition between fluid and hexatic, tetratic, and hexatic phases, respectively, and a continuous transition from the appropriate ``x"-atic to the solid. In particular, we find that systems of hexagons display KTHNY melting. In contrast, due to symmetry incompatibility between the ordered fluid and solid, systems of pentagons and plane-filling 4-fold pentilles display a one-step first-order melting of the solid to the fluid with no intermediate phase.

\end{abstract}


\maketitle

The phase behavior of two-dimensional (2D) solids is a fundamental, long-standing problem in statistical mechanics.
Whereas three-dimensional (3D) solids characteristically exhibit first order (or discontinuous) melting transitions, 2D solids can melt by either continuous or first order melting transitions and may exhibit an intermediate, so-called ``x-atic" ordered phase that is somewhere between a fluid and a solid.
Previous studies~\cite{Guo1983,Murray1987,Sachdev1985,Kusner1994,Marcus1996,Bagchi1996,Armstrong1999,Zahn1999,Eisenmann2004,Schilling2005b,Zhao2010,Bernard2011,Zhao2012,Qi2014,Qi2015b,Gantapara2015,Walsh2015,Kapfer2015} that examine two-dimensional melting find three distinct scenarios~\cite{Strandburg1988,Nelson2002}.
One, the system can exhibit a continuous fluid-to-x-atic-to-solid transition.
The x-atic phase has quasi-long-range (power-law decay) correlations in the bond order but only short-range (exponential decay) correlations in positional order.
The hexatic phase, with six-fold bond order, is the most well-known example.
The existence of continuous fluid-to-solid transitions was predicted by the Kosterlitz-Thouless-Halperin-Nelson-Young (KTHNY) theory~\cite{0022-3719-6-7-010,Halperin1978,Young1979} and has been confirmed in experiments with electrons~\cite{Guo1983} and spherical colloids~\cite{Murray1987,Kusner1994,Zahn1999,Eisenmann2004}.
The KTHNY theory of two-step melting is based upon the behavior of topological defects in the form of dislocations and disclinations.
The theory envisions that pairs of dislocations unbind to drive the transition from solid to hexatic, and then pairs of disclinations unbind to drive the transition from hexatic to fluid.
Two, the system can exhibit a first-order fluid-to-solid (or solid-to-fluid) transition, with no intervening phase.
Both this and the first scenario were realized in a system of charged polystyrene microspheres, depending on the particle diameter, which was postulated to have an effect on defect core energies~\cite{Armstrong1999}.
Three, the system can exhibit a first-order fluid-to-x-atic and a subsequent continuous x-atic-to-solid transition.
This combination of transitions is intermediate to the one-step fluid-to-solid first order transition and the two-step continuous KTHNY behavior.
It was first experimentally observed in neutral micron-scale colloidal spheres~\cite{Marcus1996}, and has been observed recently in simulations of hard disks in two dimensions~\cite{Bernard2011,Engel2013} and under quasi-2D confinement of hard spheres where out-of-plane fluctuations are limited~\cite{Qi2014}.

All three melting scenarios have been observed in experimental studies of different systems, with a variety of long and short range interactions.
Recent simulation work~\cite{Kapfer2015} finds two of the three scenarios: point particles with hard core repulsion interactions follow the third scenario and softer potentials lead to continuous melting.
In this paper, we report the occurrence of all three distinct melting scenarios in a single family of hard, regular polygons.
Hard polygons have a rotational degree of freedom, which creates the possibility for more complex entropic forces and more diverse solid phases than observed for hard disks.
By varying the number of polygon edges, we show that the melting transition scenario for a system of any given polygon is determined by the anisotropy of emergent entropic interactions along with the symmetry of the particles relative to that of the solid phase.
In particular, we show that systems of hexagons are a perfect realization of the KTHNY theory, exhibiting melting from the solid to the hexatic phase with an increase in the dislocation density, then from the hexatic to the fluid with an increase in the disclination density.
We find that systems of triangles and squares also show a continuous KTHNY-type melting transition, while systems of pentagons and 4-fold pentilles have a first order melting transition that occurs in a single step.
Finally, we show that systems of regular polygons with $n \ge 7$ behave like disks with a first order fluid-to-hexatic and continuous KTHNY-type hexatic-to-solid transition.

We focus our study on hard, convex, regular polygons because we aim to discover general unifying principles of 2D melting by filling in gaps in existing literature: regular triangles~\cite{Benedict2004,Gantapara2015,Zhao2012}, squares~\cite{Wojciechowski2004,Zhao2010,Walsh2015}, pentagons~\cite{Sachdev1985,Schilling2005b}, and heptagons~\cite{Schilling2005b} have been previously studied by both experiment and simulation.
These studies were instrumental in identifying and characterizing possible intermediate phases (triatic, tetratic, and hexatic).
We present new results for regular hexagons, octagons, \emph{etc.} up to 14-gons and clarify the results of previous simulations of hard polygons using very large simulations to conclusively determine the orders of the various melting transitions, where previous studies were too small to be conclusive.
Section II includes detailed comparisons between our results and previous simulation and experimental results.

We demonstrate that changing only the number of edges on convex regular polygons is sufficient to generate a rich array of different melting behavior, including all three known 2D melting scenarios.
We leave for future studies investigation of, e.g. rounding of polygons, where experiment~\cite{Zhao2012} and simulation~\cite{Avendano2012} have revealed additional phases.

\section{Methods}
\label{sec:methods}

\begin{figure}
\centering
\includegraphics[width=\columnwidth]{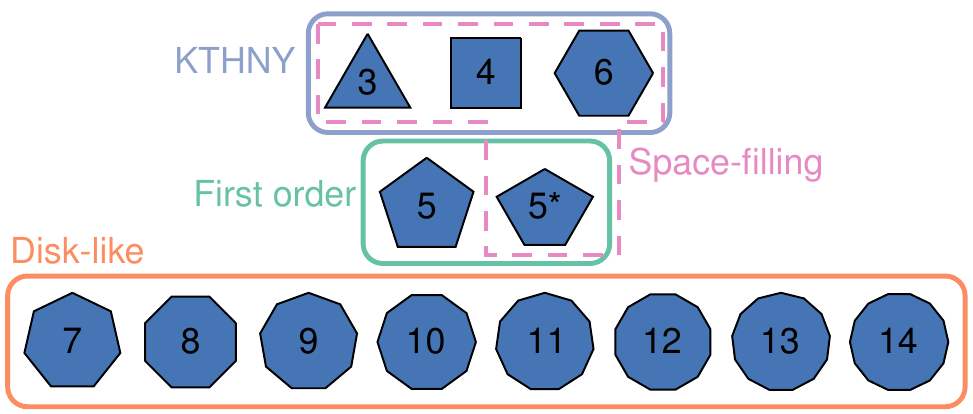}
\caption{\label{fig:shapegallery} The polygons studied in this work are the regular $n$-gons with $3\le n\le 14$ and the 4-fold pentille labelled $5^*$.}
\end{figure}

We investigate large systems of $N$ identical polygons with $n$ edges that interact solely through excluded volume interactions in a box of area $A_\text{box}$.
Particle $a$ has position $\vec{r}_a$ and orientation angle $\theta_a$.
The circumcircle diameter of the polygons is denoted as $\sigma$.
The majority of our work focuses on regular polygons ($n$-gons) with the area of a single particle $A=\sigma^2 n/2 \sin(2\pi/n)$.
We also include in our study, the 4-fold pentille~\cite{Conway2008}, which is the Voronoi cell of the Cairo pentagon pattern and thus tiles space (see figure S16 in the Supplementary Information for the tiling configuration).
\autoref{fig:shapegallery} shows all thirteen polygons and summarizes their melting behavior.

\subsection{Computational strategy}

We executed Monte Carlo (MC) simulations as large as $1024^2$ particles to obtain high precision equations of state and sample long range correlation functions to conclusively identify hexatic and solid phases, which recent hard disk simulations~\cite{Bernard2011,Engel2013} found necessary.
The MC simulations were performed in the isochoric (constant area) ensemble using the HPMC~\cite{Anderson2015a} module of HOOMD-blue~\cite{Anderson2008a,Glaser2014c,HOOMDWeb}.
HPMC (for hard particle Monte Carlo) is a parallel simulation implementation on many CPUs or GPUs using MPI domain decomposition.

Each simulation begins with the particles placed in a square, periodic simulation box.
We perform simulations for a long enough time to reach and statistically sample thermodynamic equilibrium, see Supplementary Information section II for a complete simulation protocol description, and figure S14 for an example.
Typical simulation runs initially form many domains in the system, which coalesce together over a long equilibration period of several hundred million trial moves per particle.

Occasionally, two infinite, twinned domains form in the system.
Such configurations are metastable, they are at a higher pressure than a corresponding single domain and the larger domain grows very slowly into the smaller one as the simulation progresses.
We remove stuck simulations and rerun them with different random number seeds until we obtain a cleanly equilibrated single domain sample, except in a few cases where multiple attempts to do so failed (e.g. $\phi=0.714$ in figure S10).

This work utilized significant computational resources on XSEDE~\cite{Towns2014} Stampede (222,000 SUs), OLCF Eos, OLCF Titan (115 million Titan-core-hours), and the University of Michigan Flux cluster (100,000 GPU hours).

\subsection{Equation of state}

We compute the equation of state $P^*(\phi)$ using volume perturbation techniques~\cite{Eppenga1984,Brumby2011} to measure pressure in isochoric simulations.
We report pressure in reduced units $P^* = P\sigma^2 / k_\mathrm{B}T$.
In addition to the system density $\phi=NA/A_\text{box}$, we determine the averaged local density field on a grid,
\begin{equation}
\Phi(\vec{r}_i) = \frac{A \sum_{a=1}^N H(r_c - |\vec{r}_a - \vec{r}_i|) }{\pi \cdot r_c^2},
\end{equation}
where we choose the cut-off $r_c = 20\sigma$ and $H$ is the Heaviside step function.

With the fluid density $\phi_f$ and the density of the solid (or hexatic) phase $\phi_s$ the latent heat $L=P\Delta A_\text{box}$ of a first-order phase transition is
\begin{equation}
\frac{L}{Nk_\mathrm{B}T} = P^*_c \frac{A}{\sigma^2} \left( \frac{1}{\phi_f} - \frac{1}{\phi_s} \right),
\end{equation}
where $P^*_c$ is the coexistence pressure estimated by Maxwell construction.

\subsection{Order parameters}

\begin{figure}
\centering
\includegraphics{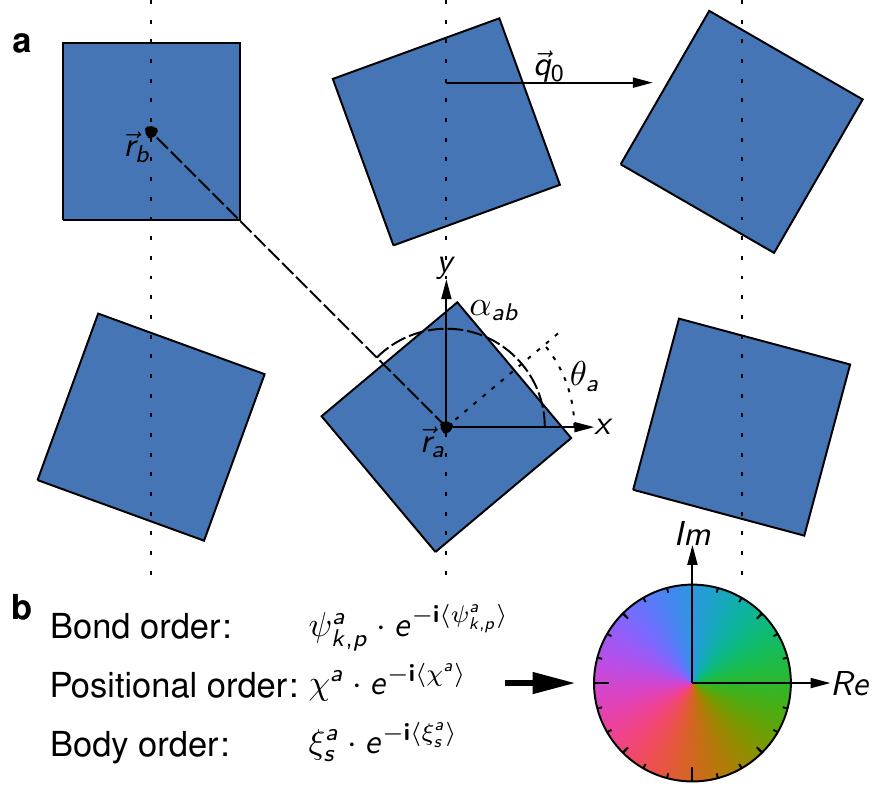}
\caption{\label{fig:analysis} a) Pictorial definitions of $\vec{r}_a$, $\vec{r}_b$, $\theta_a$, $\alpha_{ab}$, and $\vec{q}_0$ b) We map order parameters to a color wheel for visualization. Each order parameter is rotated by its average in a given frame so that the average order parameter is colored green. The color wheel is the color part of the cubehelix~\cite{Green2011} color map at constant apparent luminance ($v=0.5$, $\gamma=1$, $s=4.0$, $r=1$, $h=1$).}
\end{figure}

We use three order parameters that were previously effective at identifying the hexatic phase in the hard disk system~\cite{Bernard2011,Engel2013}.
Each order parameter is a complex number on the unit circle.
We visualize order parameter fields directly in the $x$-$y$-plane of the system by mapping the complex values of an order parameter to a color wheel (see \autoref{fig:analysis}).
Short-range order shows up as colors rapidly cycling through the color wheel, quasi-long-range order appears as patches, long-range order appears as a single solid color across the box, and two-phase regions show two separate behaviors in a single simulation box.
Independent simulation runs result in different system orientations in the box.
We rotate the order parameters so that the average in a given frame is colored green so that images may be compared by eye.

The positional order parameter
\begin{equation}
\chi^a = e^{\mathbf{i} \vec{r}_a \cdot \vec{q}_0}
\end{equation}
identifies how well the position $\vec{r}_a$ of particle $a$ fits on a perfect lattice with reciprocal lattice vector $\vec{q}_0$, as depicted in \autoref{fig:analysis}.
When all particles have the same phases in $\chi$, they are in a perfect solid.
Defects cause $\chi$ to rotate.
We choose $\vec{q}_0$ as the brightest peak in the structure factor~\cite{Bernard2011} computed with the following procedure:
(i)~Initialize a density grid with roughly $8\times8$ pixels per particle.
(ii)~At the center of each particle, place a Gaussian on this grid with standard deviation $\frac{1}{10} \sigma$.
(iii)~Take the fast Fourier transform (FFT) of the density grid to get the discretized $S(\vec{q})$.
(iv)~Smooth with a Guassian filter, standard deviation 2 pixels $S(\vec{q}) \rightarrow S_\text{smooth}(\vec{q})$.
(v)~Choose $\vec{q}_0$ from the location of the brightest pixel in $S_\text{smooth}(\vec{q})$.

The bond orientation order parameter for $k$-fold rotational symmetry
\begin{equation}
\psi^{a}_{k,p} = \frac{1}{p} \sum_{b \in \text{NN}_p(a)}e^{\mathbf{i} k \alpha_{ab}}
\end{equation}
identifies the orientation of the $p$ nearest neighbors around particle $a$.
Here $\alpha_{ab}$ is the angle the separation vector $\vec{r}_b - \vec{r}_a$ makes with respect to the positive $x$-axis, and $\text{NN}_p(a)$ is the set of $p$ nearest neighbors of $a$, see \autoref{fig:analysis} for a graphical definition.
We omit $p$ in the subscript when it is equal to $k$ and write $\psi^{a}_{k} = \psi^{a}_{k,k}$.
Some authors suggest using a morphometric approach~\cite{Mickel2013} to compute the bond orientation order parameter, which requires computing a Voronoi diagram of the system of particles.
We do not adopt this scheme because we find the use of $p$ fixed neighbors sufficient as it generates order parameter fields fully consistent with the defects present in the system.

The body orientation order parameter
\begin{equation}
\xi^{a}_{s} = e^{\mathbf{i} s \theta_a}
\end{equation}
identifies the orientation of a particle accounting for $s$-fold symmetry.
$\theta_a$ is the angle that rotates particle $a$ from a reference frame into a global coordinate system (see \autoref{fig:analysis}).
It allows us to analyze the presence of rotator phases in which $\xi^{a}_{s}$ decays to zero rapidly as a function of the separation distance.

\subsection{Correlation functions}

Correlation functions measure the behavior of the order parameters as the separation $r_{ba} = |\vec{r}_b - \vec{r}_a|$ between a pair of particles increases.
The correlation functions for bond orientation order and body orientation order are implemented in our analysis code as
\begin{equation}
C_v(r) = \frac{\sum_{b \in S(m,N)} \sum_{a=1}^N v^a (v^b)^* \delta(r - r_{ba})}{\sum_{b \in S(m,N)} \sum_{a=1}^N \delta(r - r_{ba})},
\end{equation}
where $v=\psi_{k,p}$ or $\xi_n$ and $v^*$ is the complex conjugate.
The sampling $S(m,N)$ randomly selects $m$ particles out of $N$ without replacement.

One can compute a correlation function from the positional order parameter $\chi$, but it is extremely sensitive to the choice of $\vec{q}_0$.
We observe that peaks misidentified by only a single pixel result in an apparent lack of quasi-long-range positional order.
Instead, we follow Ref.~\cite{Bernard2011} and compute positional correlation functions from
\begin{equation}
C_g(r) = \frac{ \langle g_\psi((r,0),\alpha) \rangle - 1}{\max( \langle g_\psi((r,0),\alpha) \rangle - 1)},
\end{equation}
which oscillates, so we show only identified peaks.
The signal then decays to about $10^{-3}$ at large $r$.
The function $g_\psi(\vec{r},\alpha)$ is the discretized 2D pair correlation function obtained by correlated averaging over individual measurements with index $i$,
\begin{equation}
g_\psi(\vec{r}_i, \alpha) = \frac{\sum_{b \in S(m,N)} \sum_{a=1}^N d_\kappa(R(-\alpha)\vec{r}_{ba} - \vec{r}_i)}{m / A},
\end{equation}
where $R(\beta)$ is the rotation matrix that rotates a vector by the angle $\beta$, $\Delta r$ is the bin size, and $d_\kappa$ is the coarse-grained delta function
\begin{equation}
d_\kappa(\vec{r}) = \left\{ \begin{array}{cl}
(\Delta r)^{-2} & \text{if } 0 \le r_x,r_y < \Delta r,\\
0               & \text{otherwise.}
\end{array}\right.
\end{equation}
Since the system rotates from frame to frame, it must be aligned to the bond orientation order parameter before averaging.
To do this, we compute $g_\psi(\vec{r}_i, \alpha)$ over many frames with large $m$.
We align each frame using the bond orientation order parameter averaged over all particles in the frame, $\alpha = \arg(\langle \psi_{k,p} \rangle)$.
Averaging the separately aligned $g_\psi(\vec{r}_i, \alpha)$ and then computing $C_g(r)$ significantly reduces noise~\cite{Engel2013}.

\begin{figure*}[p]
\centering
\includegraphics[width=\textwidth]{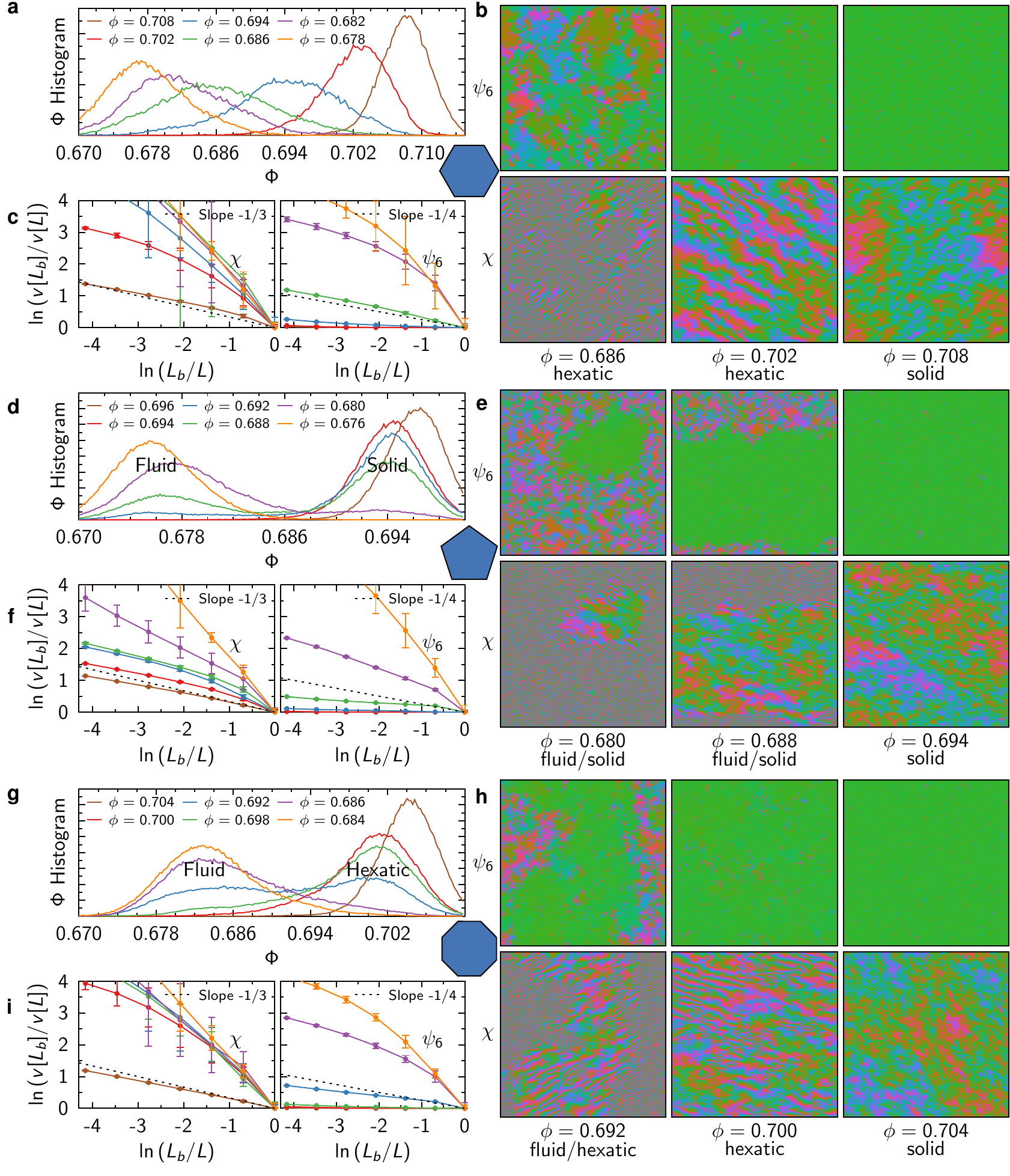}
\caption{\label{fig:phase_ident} Example phase transitions from the three melting scenarios.
We show data for (a-c)~hexagons ($N=512^2$) following the KTHNY scenario, (d-f)~pentagons ($N=1024^2$) exhibiting a first-order fluid-to-solid transition, and (g-i)~octagons ($N=1024^2$) following an intermediate scenario exhibiting a first-order fluid-to-hexatic transition.
For each shape, the top left panels show local density histograms~(a,d,g).
The right panels show bond and positional order parameters at selected densities~(b,e,h).
The bottom left panels show sub-block scaling analysis~(c,f,i).}
\end{figure*}

\subsection{Sub-block scaling analysis}

The sub-block scaling of $\chi$ is a sensitive measure of the density of the hexatic-to-solid transition~\cite{Bagchi1996,Qi2015b}.
The first density that sits on or under a line of slope $-1/3$ is at the hexatic-to-solid transition.
Similarly, sub-block scaling in $\psi$ determines the density at which the hexatic melts into the fluid, with a line of slope $-1/4$.

We perform a sub-block scaling analysis on the positional order parameter $\chi$ and the bond orientational order parameter $\psi^{a}_{k,p}$.
For this analysis the simulation box of length $L$ is divided into squares of side length $L_b$.
Within each box, we calculate
\begin{equation}
v(L_b) = \left\langle \left| \frac{1}{N} \sum_{a=1}^{N_{L_b}} v^a \right| \right\rangle,
\end{equation}
where $v=\psi_{k,p}$ or $\chi$, $N_{L_b}$ is the number of particles in a sub-block and $\langle \, \rangle$ denotes averaging over sub-blocks in the same snapshot.
Standard errors are calculated by averaging the values of $v(L_b)$ over independent frames.

\subsection{Topological defect analysis via cell-edge counting}

We generate statistics on topological defects using a Voronoi tessellation of the set of particle centers to count edges of Voronoi cells.
Each particle $a$ is assigned the number of adjacent Voronoi cells $n_a$ and the disclination charge $q_a=n_a-6$.
In the presence of well-separated topological defects, disclinations can be found by identifying particles with non-zero disclination charge, and dislocations correspond to bounds pairs of a five-coordinated particle ($q=-1$) and a seven-coordinated particle ($q=+1$).
However, while disclination charges with $|q|>1$ are very rare in our hard particle systems, particles with non-zero disclination charge are often not clearly separated or bound into pairs but instead agglomerate into larger clusters.
This makes it ambiguous to identify the locations of individual disclinations and dislocations.

To overcome this ambiguity, we cluster defects if they are in adjacent Voronoi cells and calculate the total disclination charge $q=\sum q_a$ of the cluster~\cite{Ashby1978}.
For disclination-neutral clusters ($q=0$), the total Burgers vector $\vec{b}$ of the cluster can be determined from the disclination charges and their positions,
\begin{equation}
\vec{b} = \hat{z} \times \sum_{a} q_a \vec{r}_a,
\end{equation}
where $\hat{z}$ is the unit vector along the out-of-plane axis and the cross product is performed in 3D space.
Because Burgers vectors are topologically restricted to be lattice vectors, the vector we obtain from this computation is then snapped to its closest lattice point in a hexagonal lattice whose lattice spacing is that of the ideal lattice at the density of the simulation frame.
We count the number of particles in three types of defect clusters, overall neutral ($q=0$, $\vec{b}=0$), Burgers-charged ($q=0$, $\vec{b}\neq0$), and disclination-charged ($q\neq0$) as a function of density.

\subsection{Phase determination}

We identify first-order transitions using the following criteria: (i)~a two-phase region is evident in isochoric simulations at large $N$, (ii)~the two phases have different densities, and (iii)~the equation of state has a Mayer-Wood loop~\cite{Bernard2011,Engel2013} which decreases in height as $N$ increases.
In contrast, continuous transitions have a monotonic equation of state and only a single phase at a single density is present in each frame across the entire transition.
The signature of two phase regions includes a bimodal local density histogram along with two different correlation length behaviors seen in the $\psi$ and $\chi$ order parameter fields coincident with the low and high local densities.
In all cases where we find a Mayer-Wood loop, we observe two-phase regions, and in all cases where we find a continuous equation of state, we observe only single phases across the transition.

Correlation lengths vary significantly in the simulations.
In the fluid phase, positional order $\chi$ decorrelates instantly with correlation lengths of only $1 \sigma$.
Bond orientational order $\psi_{k,p}$ persists a bit further with correlation lengths of tens of $\sigma$ but still exhibits clear short-range order.
Previous authors have used the dynamic Lindemann criterion to locate melting transition points~\cite{Zheng1998,Zahn1999}.
We focus on the correlation functions here because of their utility in discriminating hexatic and solid phases, which the dynamic Lindemann criterion is unable to do.
In the hexatic phase, $\chi$ oscillates visibly through the full color wheel along a given direction forming stripes (Fig.~3b), a behavior indicative of short-range order.
Unbound dislocations exist at the end of each stripe.
For continuous phase transitions, the oscillation period gets larger as density increases.
Solid phases lead to patchy motifs in $\chi$ (quasi-long-range order).
Two-phase regions show a combination of two of these motifs in a single system.

Our observations confirm in all cases that we have run simulations sufficiently large to allow the fluctuations in the order parameter fields of the fluids and x-atics to occur several times across the box.
Only in the solid phase regions are the order parameter fields essentially constant and vary by less than one period.

\section{Results and Discussion}

We performed identical analyses for each of the thirteen polygons investigated in this work.
Within the main text we include representative plots and snapshots to illustrate examples and explanations of the three melting scenarios.
The Supplementary Information contains detailed plots for all of the shapes in figures S1--S13, including the equation of state, local density histogram, sub-block scaling analysis, correlation functions, snapshots colored by all of the order parameters and structure factors.

\subsection{KTHNY behavior}

Our data show that systems of triangles, squares, and hexagons have continuous fluid-to-x-atic and continuous x-atic-to-solid transitions.
We refer to the monotonic equations of state in Figure S1, Figure S2, and Figure S5 for the triangle, square, and hexagon data, respectively.
Triangles and hexagons have hexatic order with a signal in $\psi_6$, and squares have tetratic order with a signal in $\psi_4$.
\autoref{fig:phase_ident}a--c demonstrates the continuous transitions for hexagons, with a single density at each state point and gradually developing order in the order parameter fields.

\subsubsection{Decay of correlation functions}

The first state point showing quasi-long-range bond orientation order in the hexagon system is $\phi=0.686$.
At this density, positional order is extremely short-ranged, persisting for only a few particle diameters.
As density increases, the decay length of the positional order becomes longer, up to hundreds of particle diameters at $\phi=0.702$, and begins to diverge.
At $\phi=0.708$, positional order switches to quasi-long-range and the system is in the solid phase as determined by sub-block scaling.
KTHNY theory predicts a slope of $-1/4$ in $\psi_6$ scaling at the fluid-to-hexatic transition, a perfect match for the $\phi=0.686$ line in \autoref{fig:phase_ident}c.
The theory also predicts a slope of $-1/3$ for $\chi$ scaling, which similarly matches the scaling for $\phi=0.708$.
At the same density, the directly computed positional correlation $C_g(r)$ length is beginning to diverge so it is difficult to determine the exact density of the hexatic-to-solid transition from correlation functions alone.

\subsubsection{Topological defects and local order}

The KTHNY theory envisions a picture of tightly bound defects transitioning to free dislocations at the solid-hexatic transition, and to free disclinations at the hexatic-fluid transition.
Interestingly, both Bernard \emph{et al.}~\cite{Bernard2011} and this work show that defects are not free, but rather form large and complex clusters, often highly anisotropic, indicating medium-ranged effective inter-defect interactions. In the extreme case that defects cluster into one-dimensional strings, such a scenario would lead to grain-boundary induced melting~\cite{Fisher1979,Chui1983}.

\begin{figure*}
\centering
\includegraphics[width=\textwidth]{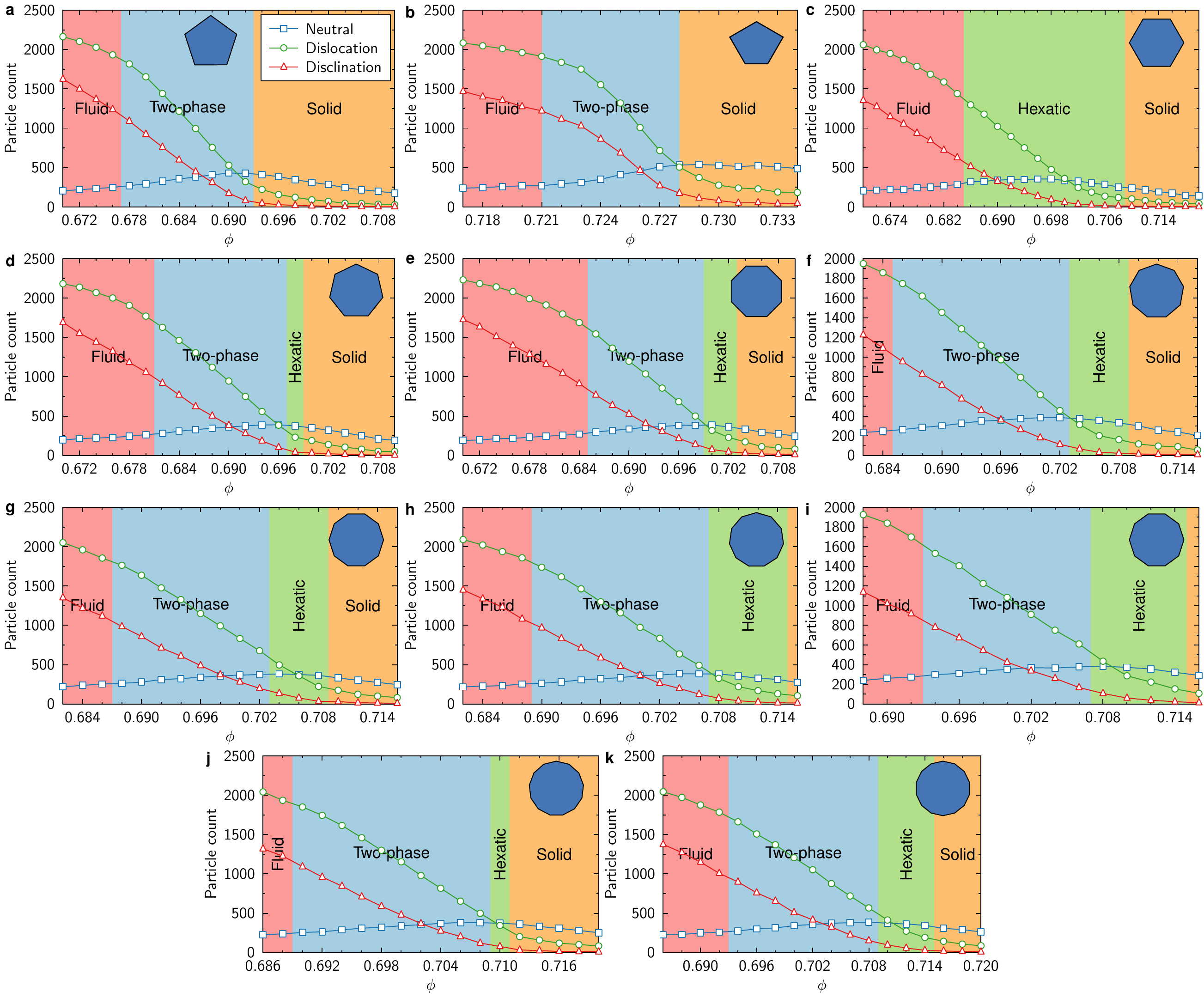}
\caption{\label{fig:defects} Defect counts as a function of density obtained with cell-edge counting.
The particle count is the number of particles that belong to any defect cluster classified into neutral, Burgers-charged and disclination-charged clusters.
The figures show results of the defect analysis plots for polygons $n=5,6,7,8,12$, averaged over 4--8 runs with 19--49 frames per run at the system size $N=128^2$.
The defect count algorithm is expensive, so we do not run it on the largest size simulations we performed.
We also show the phase diagram overlaid on the counts.
Error bars at one standard deviation are smaller than the symbol size.
}
\end{figure*}

\autoref{fig:defects}c shows the density dependence of the number of particles comprising clusters of defects that are overall neutral, those with a Burgers charge, and those with disclination charge. A large number of free dislocation clusters with non-zero Burgers vectors simultaneous with few free disclinations is evidence for the hexatic phase.
We observe that the count of particles in Burgers-charged clusters begins increasing just below $\phi=0.710$ at the solid-to-hexatic transition.
The count of particles in disclination-charged clusters remains low, and starts ramping up slowly in the middle of the hexatic phase at $\phi=0.696$.
Throughout the hexatic phase, we find many more particles in Burgers-charged clusters than in disclination-charged clusters.
This is consistent with the two-step KTHNY melting scenario and the continuous phase transition we observe for hexagons.

\begin{figure}
\centering
\includegraphics[width=7cm]{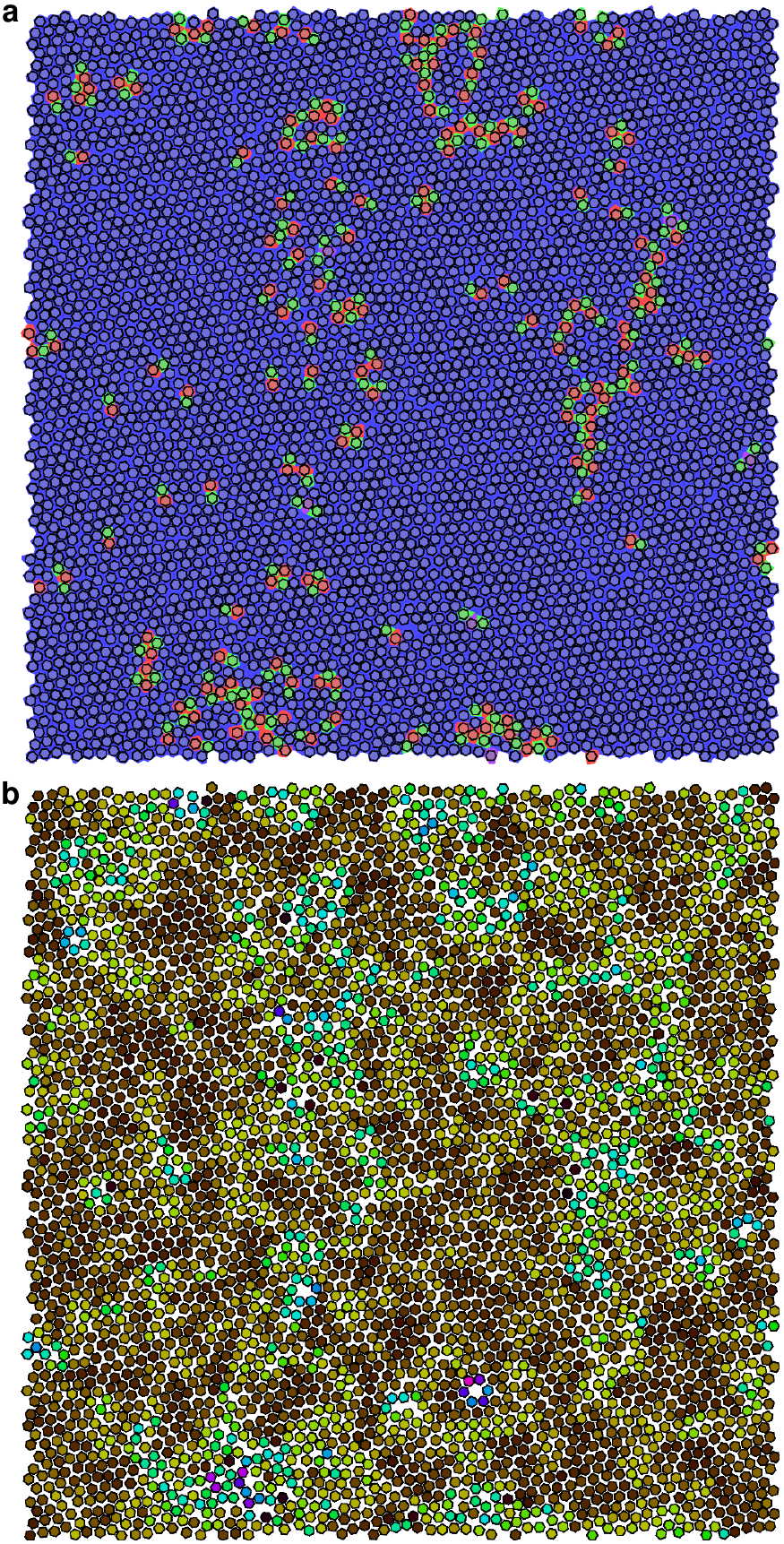}
\caption{\label{fig:defect-configuration} Representative snapshot of the hexagon system $N=128^2$ at density $\phi=0.690$.
(a)~Particles are colored by the number of edges in the Voronoi cell.
Red, blue, and green cells have seven, six, and five sides respectively.
(b)~Particles are colored by $\langle IQ \rangle-IQ$.
Blue indicates large deviation from $\langle IQ \rangle$ and yellow indicates little deviation.}
\end{figure}

\autoref{fig:defect-configuration}a shows the structure of the defects in a system of hexagons at $\phi=0.690$, in the hexatic phase.
The density of free dislocations (red and green pairs) is much higher than that of free disclinations (there is a single lone green particle in the image), but the structure is more complicated because the defects combine into large clusters.
The left panel of Supplementary Movie 1 shows how defects migrate as the Monte Carlo simulation progresses via local moves for a simulation of hexagons at $\phi=0.690$.
Dislocations are unstable and highly mobile, quickly hopping between sites and popping in and out of existence.
We find that the count of nearest neighbors is very sensitive to small changes in particle coordinates.

We also compute the isoperimetric quotient $IQ=4\pi A/P^2$ for each Voronoi cell, where $A$ is the area and $P$ is the perimeter of the cell.
Regular polygons have the value $IQ=\pi/\left(n\tan\frac{\pi}{n}\right)$.
More elongated Voronoi cells have low $IQ$ and indicate locations of large crystal deformity, which makes $IQ$ an approximate indicator of defects.
\autoref{fig:defect-configuration}b and the right panel of Supplementary Movie 1 color particles by $\langle IQ \rangle-IQ$.
Blue indicates large deviation from $\langle IQ \rangle$ and yellow indicates little deviation.
If $IQ > \langle IQ \rangle$, the particle is darkened, as these particles are closest to having regular crystal environments.

The coloring scheme based on the isoperimetric quotient $IQ$ is a continuous quantity and, as seen in the movie, is less sensitive to small particle displacements.
In contrast, the number of edges of a Voronoi cell changes discontinuously across Monte Carlo steps.
This suggests $\langle IQ \rangle - IQ$ might be a more robust indicator of defect concentrations on short time scales, while cell-edge counting is simple and reliable for long time averages over uncorrelated frames.
We also note that in systems of triangles and squares, defects are even more delocalized and Voronoi cell-edge counting is unable to identify defects, while $\langle IQ \rangle - IQ$ can still identify areas where defects are present.

\subsubsection{Particle alignment}

At all densities we simulate for triangles, squares, and hexagons, the body order parameter $\xi_s$ closely matches that of the bond order parameter $\psi_{k,p}$.
We quantify this by computing the cross-correlation between the two quantities $|\langle \psi_{k,p}\xi_{s}^* \rangle|$ in the fluid, hexatic, and solid phases (\autoref{tab:cross_correlation}).
We expect high bond-body cross-correlation at high density when particle rotation becomes locked in, as found in previous studies~\cite{Schilling2005b}.
However, a strong signal is present even in the fluid for triangles, squares, and hexagons.

\begin{table}
\tabcolsep1.5mm \centering
\begin{tabular}{lcccc}
\hline\hline
& & \multicolumn{3}{c}{Phase} \\
Polygon & Cross-correlation & Fluid & x-atic & Solid \\
\hline
Triangle & $|\langle \psi_{6,3}\xi_6^* \rangle|$ & 0.117 & 0.221 & 0.236 \\
Square & $|\langle \psi_{4}\xi_4^* \rangle|$ & 0.449 & 0.790 & 0.810 \\
Pentagon & $|\langle \psi_{6}\xi_{10}^* \rangle|$ & 0.001 & - &  0.001 \\
Hexagon & $|\langle \psi_{6}\xi_6^* \rangle|$ & 0.244 & 0.294 &  0.299 \\
Heptagon & $|\langle \psi_{6}\xi_{14}^* \rangle|$ & 0.001 & 0.001 &  0.001 \\
Octagon & $|\langle \psi_{6}\xi_{8}^* \rangle|$ & 0.001 & 0.001 &  0.001 \\
Dodecagon & $|\langle \psi_{6}\xi_{12}^* \rangle|$ & 0.001 & 0.039 &  0.039 \\
\hline\hline
\end{tabular}
\caption{\label{tab:cross_correlation} Cross-correlation between bond and body orientation in the fluid, x-atic, and solid phases. We select the highest density pure fluid phase, the highest density pure x-atic phase, and the lowest density pure solid phase. Errors of two standard deviations of the mean are approximately 0.001 for all values.}
\end{table}

\begin{figure*}
\centering
\includegraphics[width=0.7\textwidth]{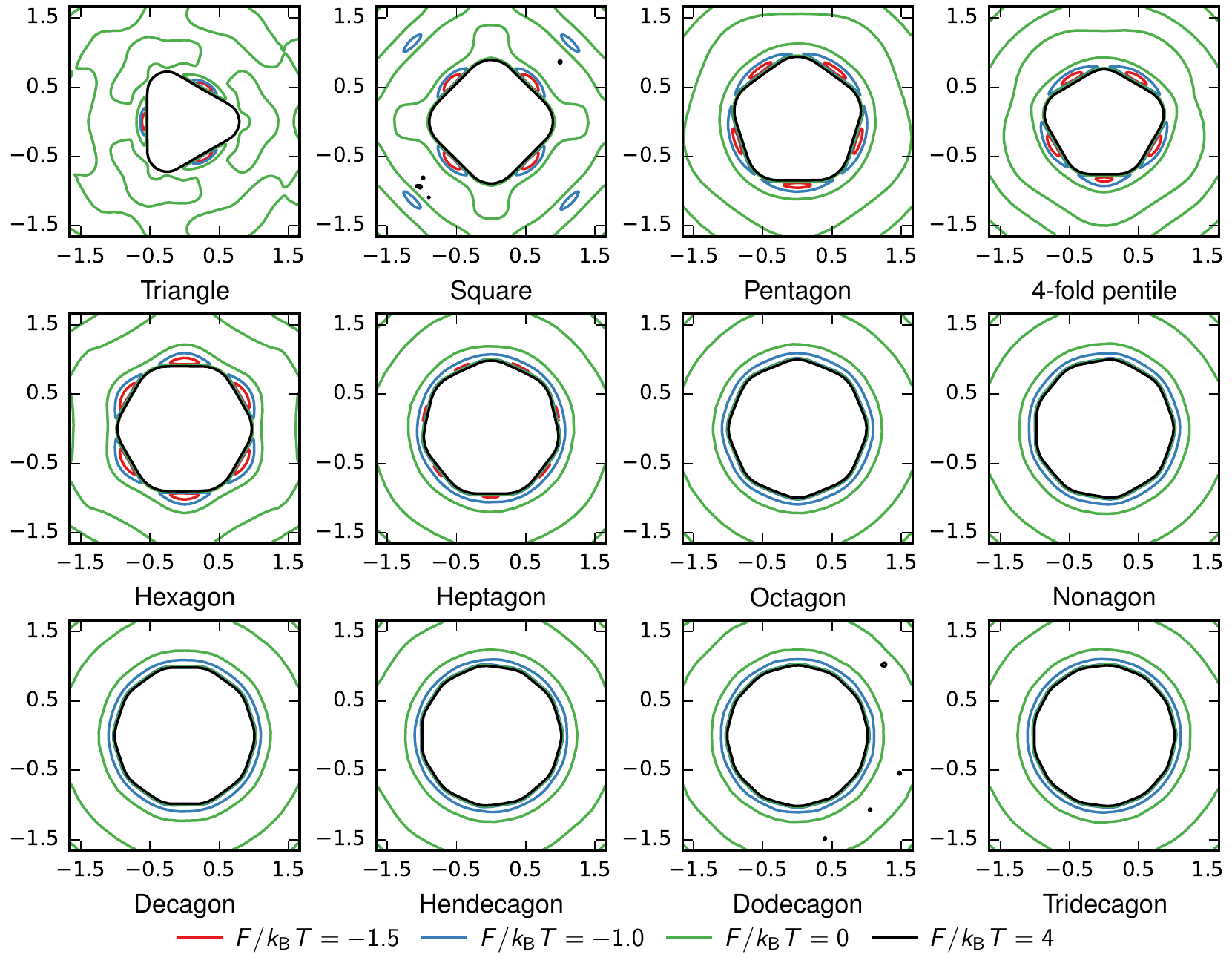}
\caption{\label{fig:pmft} Potential of mean force and torque plots for polygons $n$=3--13 computed at the highest density fluid for each polygon.
We choose the zero energy reference as the average over a large region $25\sigma$ from the center (not shown).
The data are shown as a contour plot with selected free energy or entropy well contours. Computed from $N=256^2$ simulations.}
\end{figure*}

The cross-correlation results demonstrate alignment of the edges of neighboring particles.
This can be interpreted as an effect of directional entropic forces~\cite{Damasceno2012d}.
We quantify the alignment with the potential of mean force and torque (PMFT)~\cite{VanAnders2014e}, computed from runs in the highest density pure fluid phase for each polygon.
For each particle, we determine the relative position of its neighbors and construct histograms using many frames to obtain reliable averages.
The free energy $F$ of a given configuration is related to the negative logarithm of the histogram.
We average an area in the PMFT $25\sigma$ from the center to set $F=0$ as a baseline.

\autoref{fig:pmft} shows the results of this calculation.
All low-symmetry regular polygons ($n < 7$) have distinctly separated wells for edge-to-edge contacts at $F=-1.5 k_\mathrm{B} T$.
This indicates preferential edge-to-edge alignment in the solid by strong directional entropic forces.
But only when the symmetry of the particle shape is compatible with the symmetry of the bond order, as found in hexagons, triangles, and squares, does such edge-to-edge alignment promote local solid motifs in the fluid.
Such a pre-ordered fluid allows the correlation length to increase smoothly from short-range to quasi-long-range and the transition to the solid to be continuous.

\begin{figure}
\centering
\includegraphics[width=\columnwidth]{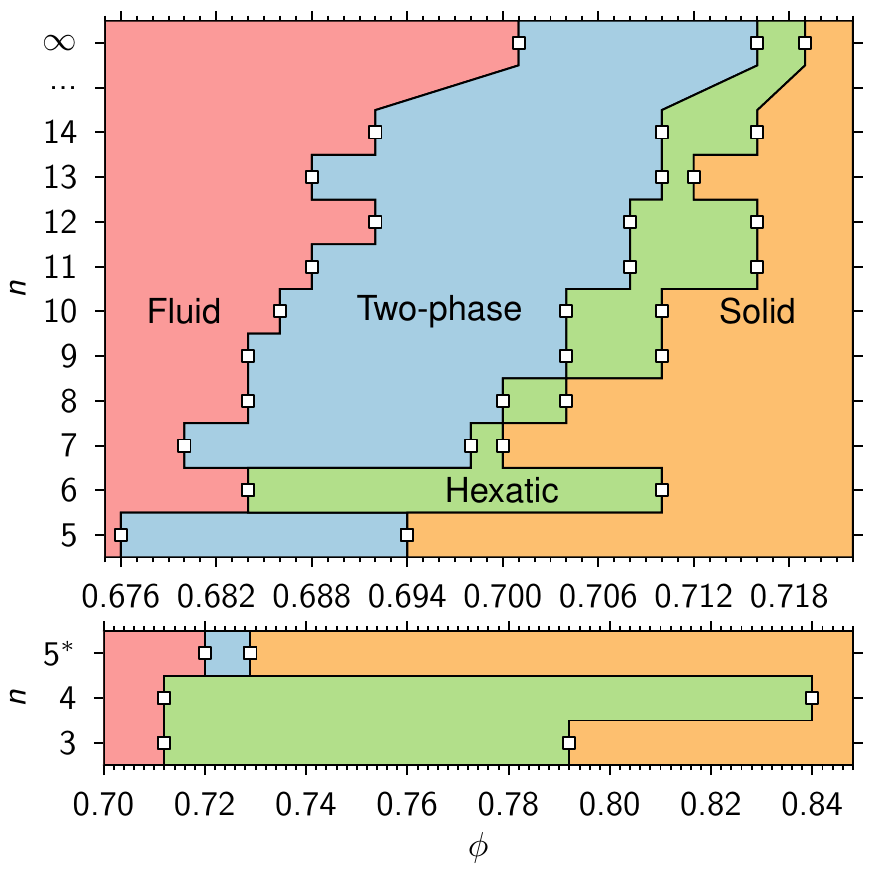}
\caption{\label{fig:phase_diagram} Phase diagram of hard polygon melting behavior. Disk results ($n \rightarrow \infty$) are from~\cite{Bernard2011,Engel2013}.
The label $5^*$ refers to the 4-fold pentille.
The $n=3$ solid is a honeycomb lattice with alternating triangle orientations, the $n=4$ the solid is a square lattice, and the $n \ge 5$ solids are all hexagonal.}
\end{figure}

Triangles and squares have a distinct overall behavior from hexagons due to the delocalized defects.
They have a smeared out phase transition between $\phi=0.71$ and $\phi \approx 0.8$ (\autoref{fig:phase_diagram}).
Their equations of state are smooth and slowly increasing with barely detectable kinks (Figures S1 and S2).
In comparison, the hexagon equation of state has sharp kinks and almost levels off through the transition (Figure S5).
Interestingly the melting transition of hexagons becomes more evidently continuous as the system size increases and could, in fact, be mistaken for a first order transition in small systems of only a few thousand particles.

Our data agree qualitatively with previous studies. Simulations of equilateral triangles using an approximate event-driven molecular dynamics method showed a continuous fluid-to-liquid crystal-like phase transition~\cite{Benedict2004}. Monte Carlo simulations revisiting that system reported a fluid-to-hexatic transition at $\phi=0.70$ and a hexatic-to-solid transition at $\phi=0.87$~\cite{Gantapara2015}.
Ref.~\cite{Gantapara2015} finds a chiral phase at $\phi \ge 0.89$, which we do not observe as we focus on the melting behavior at lower density $\phi < 0.80$.
Although no experiments have yet been reported on hard, mathematically regular triangles, rounded triangular colloidal platelets exhibit a fluid-to-hexatic transition, but the order of the transition was not determined~\cite{Zhao2012}.
Ref.~\cite{Zhao2012} also finds a phase with local chiral symmetry breaking that we do not observe for hard regular triangles at $\phi < 0.85$.
Monte Carlo simulations of squares show a continuous fluid-to-tetratic transition at $\phi=0.7$~\cite{Wojciechowski2004}.
Experiments on vibrated granular squares (LEGOs) find tetratic orientational order in the range $\phi=0.70$ to $0.74$~\cite{Walsh2015}.
We are not aware of any studies on systems of hard hexagons to compare with.

\subsection{First-order fluid-to-solid}

Our data clearly show that systems of regular pentagons have a first-order transition directly from the fluid to the solid.
This is seen in \autoref{fig:phase_ident}d-f at $\phi=0.688$.
There is a stripe of low density (local density $\Phi=0.676$) next to a stripe of solid ($\Phi=0.694$) with quasi-long-range positional order.
The two phase region starts at $\phi=0.680$ and the pure solid phase starts at $\phi=0.694$.
The two phase region is coincident with a Mayer-Wood loop in the equation of state (Figure S3).
The symmetry of the pentagon is not compatible with hexagonal ordering in the solid, so even though it has strong entropic edge-edge bonds, the body-bond cross-correlation (\autoref{tab:cross_correlation}) is zero and the phase transition becomes first-order.

The one-step melting process of pentagons also shows up in the defect counts \autoref{fig:defects}a,b.
We observe sharp kinks just below the pure solid phase at $\phi=0.692$ for counts in clusters with Burgers and disclination charges.
One-step melting is supported by this coincident increase in the number of free dislocations and free disclinations.

Our results are consistent with previous studies of pentagons. Monte Carlo simulations on small systems of pentagons showed a transition from a fluid to a hexagonal rotator crystal at $\phi=0.68$~\cite{Schilling2005b}.
The same fluid to rotator crystal transition was also reported with rounded colloidal pentagons at $\phi=0.66$~\cite{Zhao2009}.
The data in that work suggests a possible hexatic phase, but is inconclusive due to small system sizes within the field of view of the camera. Our simulations indicate there should be no hexatic phase with zero rounding.
Pentagon phase behavior has also been studied in systems of 5-fold symmetric molecules~\cite{Bauert2009} and with vibrating shaking tables~\cite{Duparcmeur1995,Sachdev1985}. None of these previous studies, however, conclusively demonstrated the first order, one step nature of the melting transition in pentagons.

Of all the regular shapes we studied, only triangles, squares, and hexagons fill space.
These are also the only three KTHNY-type shapes.
To test if another space filling polygon behaves similarly, we conducted simulations of the 4-fold pentille~\cite{Conway2008}, an irregular pentagon with two different edge lengths that tiles space with 4-fold symmetry.
We find that the 4-fold pentille behaves like the regular pentagon, with a first-order fluid-to-solid transition and no intermediate hexatic, though the transition occurs at a higher pressure (Figure S4).
At high fluid densities, directional entropic forces (\autoref{fig:pmft}) are blurred by the edge lengths and the resulting ten-fold particle body order is not compatible with either the tiling or the hexagonal solid motifs, so the transition is first-order.
This suggests the space-filling property itself is not the factor triggering KTHNY melting but instead the similarity between the local order in the dense fluid and the local order in the solid.

\subsection{Disk-like behavior}

\begin{figure}
\centering
\includegraphics[width=\columnwidth]{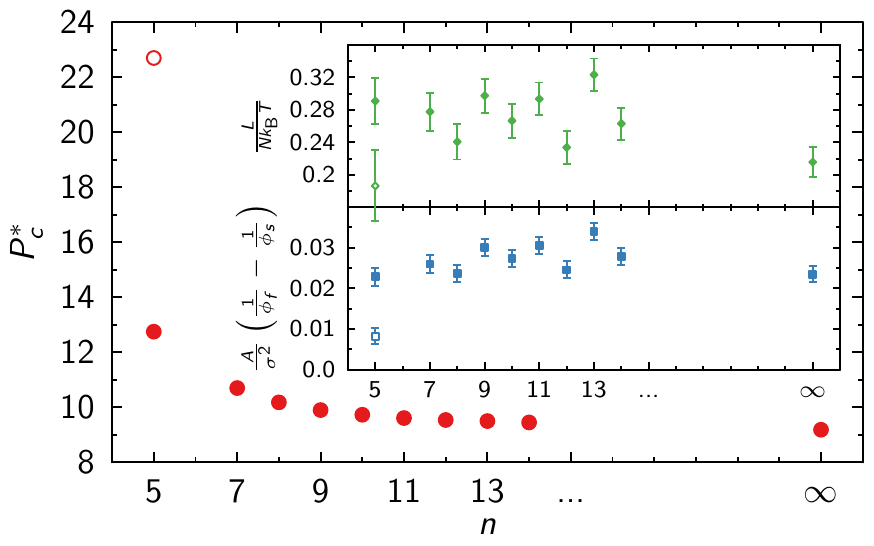}
\caption{\label{fig:latent_heat} Latent heat (green diamonds) of the first-order phase transition and the two component factors plotted separately.
$P^*_c$ (red circles) is the coexistence pressure and the area difference per particle $\delta A$ is shown below (blue squares).
Filled symbols are regular polygons, and the open symbols are the 4-fold pentille.
Error bars are estimated at 0.001 error in the transition densities.
Data for disks ($n \rightarrow \infty$) is from~\cite{Bernard2011}.}
\end{figure}

\autoref{fig:phase_diagram} summarizes the phases found for all regular polygons studied in this work and includes the hard disk phase diagram ($n \rightarrow \infty$) for comparison.
Disks have a first-order fluid-to-hexatic transition, a narrow region of stability for the hexatic phase, and a continuous hexatic-to-solid transition.
A similar smooth behavior with $n$ is found in the latent heat of the first-order transitions (\autoref{fig:latent_heat}).
The coexistence pressure decreases monotonically as $n$ increases, approaching that of disks.
The 4-fold pentille is an outlier compared to the regular polygons with almost a factor of two higher coexistence pressure.

We executed simulations up to $n=14$ in this work and find that all regular polygons with $n \ge 7$  have first-order fluid-to-hexatic and continuous hexatic-to-solid phase transitions.
\autoref{fig:phase_ident}g-i illustrates this for a system of octagons.
In the two-phase region we see a mixture of fluid and hexatic phases.
This is the same behavior as seen for disks, the only difference is that the transition shifts to lower packing fraction as $n$ decreases. This shift is expected from the increasing anisotropy and relative strength of directional entropic forces with decreasing $n$.
At $n=7$ the start of the phase transition is at $\phi=0.680$, increases to $\phi=0.692$ by $n=12,14$ then again increases to $\phi=0.702$ for disks.
We observe a small jump in critical density from $n=14$ to disks, despite the very close coexistence pressures.
We surmise that all regular polygons with $n \ge 7$ have a first-order fluid-to-hexatic transition followed by a continuous hexatic-to-solid transition, with the transition range shifting higher in density with larger $n$.

Of the studied polygons with $n \ge 7$, only the dodecagon with $n=12$ has a particle symmetry compatible with the bond order in the solid.
Although for this polygon strong directional entropic forces have the potential to drive a continuous transition from the fluid to the hexatic and solid phase, the simulations show a clear first-order transition for dodecagons.
There is no body-bond cross-correlation in the dodecagon fluid (\autoref{tab:cross_correlation}), but it does appear weakly in the hexatic phase.
The relatively short edge lengths in the dodecagon lower the strength of edge-edge alignment to the point where it is not strong enough to encourage local hexagonal motifs in the fluid, and as a result the transition from bond orientation disorder to order becomes sharp.
All regular polygons with $n \ge 7$ have smeared out $F=-1k_\mathrm{B}T$ contact wells (\autoref{fig:pmft}) that encircle the polygon, indicating weaker edge-edge alignment than triangles, squares, and hexagons, allowing a much more isotropic behavior.
This is why all regular polygons $n \ge 7$ share the same fluid-solid transition properties as disks.

Regular polygons with $n \ge 7$ exhibit an intermediate hexatic phase with a very narrow region of stability and a first-order transition to the fluid.
As a result, the defect counts (\autoref{fig:defects}d-k) for these polygons is not as clear.
The Burgers-charged and disclination-charged counts for these polygons have sharper kinks than the hexagons, but they do not parallel each other as closely as in the pentagons.
Despite this, the hexatic-to-solid transition follows KTHNY predictions.
In particular, the sub-block scaling analysis for $\chi$ predicts the hexatic-to-solid transition density with high accuracy, as shown in \autoref{fig:phase_ident}i where the scaling line for $\phi=0.704$ falls almost exactly on the dotted line predicted by theory.

\section{Conclusion}

In 1988, Strandburg wrote~\cite{Strandburg1988}: ``For a decade now the nature of the two-dimensional melting transition has remained controversial.''
At that time the three fluid-to-solid transition scenarios discussed here were considered, but in many cases the available evidence was inconclusive as to which system shows which scenario and whether all scenarios indeed occur.
2D simulations have come a long way since 1988.
With the advancement of high-performance computing we can now obtain correlation functions with high enough precision and rely on additional analysis techniques like order parameter fields, sub-block scaling, and cell-edge counting.
Together, these tools reliably identify x-atic phases and resolve the nature of the 2D melting transition.
Today, we can confidently state that the three scenarios discussed already 30 years ago indeed occur and, in fact, can be observed in a single system, namely the hard polygon family studied in the present work.

As we have demonstrated, the polygon's shape symmetry with respect to the lattice of the solid phase together determine the melting scenario.
Systems of triangles, squares, and hexagons follow the KTHNY scenario well.
They promote strong directional entropic forces that preorder the fluid into symmetries that are compatible with the solid.
Pentagons and plane-filling 4-fold pentilles have similar strong anistropic aligning forces in the fluid, but their symmetry is incompatible with the hexagonal order of their solid phase.
As a result they display a clear one-step first-order melting of the solid to the fluid with no intermediate phase.
Regular polygons with sufficiently many ($n\ge 7$) edges have no preferential alignment.
They show a close resemblance to hard disk behavior and exhibit a first order fluid-to-hexatic phase transition and a continuous hexatic-to-solid transition, which is the intermediate scenario between KTHNY and one-step first-order melting.

Our results show that some 2D particles exhibit a KTHNY melting scenario when local coupling is strong and a hard-disk melting scenario when local coupling is weak. Our findings agree with a recent study~\cite{Kapfer2015} of disks interacting via soft $r^{-m}$ potentials; for $m>6$, disks are hard enough to exhibit hard-disk melting, whereas for $m\leq 6$, they exhibit KTHNY melting. In  our system, we correlate ``softness" achieved with $n\leq 6$ with strong anisotropy in the entropic force field. For polygons with $n\leq 6$, the anisometry is sufficient to allow other particles to approach at distances well inside the corresponding circumcircle of the polygon. This ``overlap" creates the equivalent of a soft effective interaction.
Rounding polygon edges limits how close neighboring particles may approach, and e.g. systems of sufficiently rounded hexagons should demonstrate disk-like behavior -- we leave such a study for future work.
More generally, the melting transition depends strongly on the symmetry of local interactions, not just the strength, when one has anisotropic interactions. In the present study, this anisotropic coupling is provided by emergent entropic forces, but we predict that anisotropic coupling of any origin is sufficient to produce the variety of melting scenarios we have observed in polygons.

\section{Acknowledgments}

This work was supported by the National Science Foundation, Division of Materials Research Award \#DMR 1409620 (to J.A.A.\ and S.C.G.), US Army Research Office Grant W911NF-10-1-0518 (to S.C.G.\ and M.E.), National Science Foundation Graduate Research Fellowship Grant DGE 1256260 (to J.A.), and the DOD/ASD(R\&E) under Award No.\ N00244-09-1-0062.
Any opinions, findings, and conclusions or recommendations expressed in this publication are those of the authors and do not necessarily reflect the views of the DOD/ASD(R\&E).
M.E. acknowledges funding by Deutsche Forschungsgemeinschaft through the Cluster of Excellence Engineering of Advanced Materials.

All data analysis in this work was performed using Freud, an in-house Python-driven high-performance toolkit developed by the Glotzer Group.
Matthew Spellings implemented the complex correlation function routine and the cubeellipse color map.
Eric S. Harper implemented the PMFT analysis code and parallelized all modules.
We thank Wenbo Shen for suggesting that we examine correlations in the body orientation order parameter, and Greg van Anders for a critical read of the manuscript.

This work used resources of the Oak Ridge Leadership Computing Facility at the Oak Ridge National Laboratory, which is supported by the Office of Science of the U.S. Department of Energy under Contract No. DE-AC05-00OR22725, on the Extreme Science and Engineering Discovery Environment (XSEDE), which is supported by National Science Foundation grant number ACI-1053575, and also on computational resources and services provided by Advanced Research Computing Technology Services at the University of Michigan, Ann Arbor.

\section{Author Contributions}

J.\ A.\ Anderson ran the large scale simulations and analyzed the data.
J.\ Antonaglia analyzed the defects.
J.\ A.\ Millan executed small scale simulations and determined regions of interest.
J.\ A.\ Anderson and M.\ Engel conceived the project.
S.\ C.\ Glotzer supervised the research.
All authors discussed methods and results and participated in writing the manuscript.

\bibliography{manuscript}

\end{document}